\documentclass[aps,floats]{revtex4}
\usepackage{amsmath,amssymb}
\usepackage{graphicx,epsfig}
\usepackage[greek,english]{babel}

\begin{document}
\bibliographystyle {plain}

\def\oppropto{\mathop{\propto}} 
\def\opsimeq{\mathop{\simeq}}
\def\opoverderline{\mathop{\overline}}
\def\operarrow{\mathop{\longrightarrow}}
\def\opsim{\mathop{\sim}}

\def\fig#1#2{\includegraphics[height=#1]{#2}}
\def\figx#1#2{\includegraphics[width=#1]{#2}}


\title{ Strong Disorder Renormalization for the dynamics of Many-Body-Localized systems   : \\ 
 iterative elimination of the fastest degree of freedom via the Floquet expansion } 


\author{ C\'ecile Monthus }
 \affiliation{Institut de Physique Th\'{e}orique, 
Universit\'e Paris Saclay, CNRS, CEA,
91191 Gif-sur-Yvette, France}

\begin{abstract}
The Vosk-Altman Strong Disorder Renormalization for the unitary dynamics of various random quantum spin chains is reformulated as follows : the local degree of freedom characterized by the highest eigenfrequency $\Omega$ can be considered as a high-frequency-Floquet-periodic-driving for the neighboring slower degrees of freedom. Then the two first orders of the high-frequency expansion for the effective Floquet Hamiltonian can be used to generate the emergent Local Integrals of Motion (LIOMs) and to derive the renormalization rules for the effective dynamics of the remaining degrees of freedom. The flow for this  effective Floquet Hamiltonian is equivalent to the RSRG-X procedure to construct the whole set of eigenstates that generalizes the Fisher RSRG procedure constructing the ground state. This general framework is applied to the random-transverse-field XXZ spin chain in its Many-Body-Localized phase, in order to derive the renormalization rules associated to the elimination of the biggest transverse field and to the elimination of the biggest coupling respectively.

\end{abstract}

\maketitle

\section{ Introduction }

In the field of Many-Body-Localization, one is interested into the unitary dynamics of isolated quantum interacting models (see the recent reviews \cite{revue_huse,revue_altman,revue_vasseur,revue_imbrie,revue_rademaker,review_mblergo,review_prelovsek,review_rare} and references therein).
The Many-Body-Localized phase is characterized by the emergence of an extensive number of Local Integrals of Motion called LIOMs
\cite{emergent_swingle,emergent_serbyn,emergent_huse,emergent_ent,imbrie,serbyn_quench,emergent_vidal,emergent_ros,
emergent_rademaker,serbyn_powerlawent,c_emergent,ros_remanent,wortis,c_liom,counting_lioms,maj_pol}. 
Among the various approaches to identify these LIOMs, 
two closely-related generalizations of
the Strong Disorder Real-Space RG approach 
developed by Daniel Fisher \cite{fisher_AF,fisher,fisherreview} to construct the ground states of random quantum spin chains (see the review \cite{strong_review})
 have been proposed:

(i) the RSRG-X procedure has been formulated to construct the whole set of excited eigenstates 
 \cite{rsrgx,rsrgx_moore,vasseur_rsrgx,yang_rsrgx,rsrgx_bifurcation,c_mblcayley,c_rsrgxMaj}.

(ii) the RSRG-t procedure has been developed by Vosk and Altman \cite{vosk_dyn1,vosk_dyn2}
in order to construct the effective dynamics via the iterative elimination of the degree of freedom oscillating with the highest frequency.

In the present paper, we reformulate the Vosk -Altman RSRG-t procedure (ii) within the Floquet theory of periodically driven systems at high frequency (see the review \cite{floquet_review})
 in order to obtain a general recipe to compute the renormalization rules for more general models
 than the specific Hamiltonians and initial conditions considered in  \cite{vosk_dyn1,vosk_dyn2}.
This reformulation also allows to make a direct correspondance with the RSRG-X procedure (i) for the excited eigenstates.
As an example of application, we consider the random-transverse-field XXZ spin chain, which is the 
standard model in the field of Many-Body-Localization with many available numerical results
\cite{kjall,alet,alet_dyn,luitz_tail,badarson_signa,auerbach,znidaric_dephasing,prelo_dyn,znidaric_lindblad,
luitz_bimodal,garcia,luitz_anomalous,luitz_operator,luitz_information}.

The paper is organized as follows.
In section \ref{sec_gen}, the Vosk-Altman RSRG-t procedure for the dynamics is reformulated using the  
Floquet theory of periodically driven systems at high frequency.
The application to the random-transverse-field XXZ spin chain is described in the following sections :
the renormalization rules for the elimination of the biggest transverse field is derived in section \ref{sec_h},
while the renormalization rules for the elimination of the biggest coupling is derived in section \ref{sec_j},
Our conclusions are summarized in section \ref{sec_conclusion}.

\section{ General principle of the renormalization procedure for the dynamics }

\label{sec_gen}

\subsection{ Reminder on the high-frequency expansion for Floquet-periodically-driven systems }

The Floquet theory of periodically driven quantum systems at high frequency (see the review \cite{floquet_review} and references therein)
yields in particular the following result :
when a quantum system is driven with some periodic Hamiltonian of frequency $\Omega$ defined by its Fourier series
\begin{eqnarray}
H(t) = \sum_{l =-\infty}^{+\infty} {\cal H}_l e^{i l \Omega t}
\label{hfourier}
\end{eqnarray}
the effective Floquet Hamiltonian can be computed via the high-frequency expansion (see \cite{fishman1,fishman2,dal1,dal2,eck}
and section 3.2 together with Appendix A  of the the review \cite{floquet_review})
\begin{eqnarray}
H_{eff}= H_{eff}^{(0)}+  H_{eff}^{(1)} + O\left(\frac{1}{\Omega^2} \right)
\label{heff}
\end{eqnarray}
where the contribution of order $O(1)$ is simply the time-average of the Hamiltonian $H(t)$ 
as given by the Fourier coefficient $l=0$ in Eq. \ref{hfourier}
\begin{eqnarray}
 H_{eff}^{(0)} = \frac{1}{T} \int_0^T dt H(t) = {\cal H}_0
\label{heff0}
\end{eqnarray}
while the contribution of order $\frac{1}{\Omega}$ involves commutators between  Fourier coefficients
of Eq. \ref{hfourier}
\begin{eqnarray}
 H_{eff}^{(1)} = \frac{1}{\Omega} \sum_{l =1}^{+\infty} \frac{1}{l} [ {\cal H}_l, {\cal H}_{-l} ]
\label{heff1}
\end{eqnarray}
This high-frequency expansion makes sense if the frequency $\Omega$ of the driving
 is much bigger than all natural frequencies of the system.

Note that besides the high-frequency-expansion quoted in Eqs \ref{heff}, \ref{heff0}, \ref{heff1} for the effective Hamiltonian $H_{eff}$,
there also exists the corresponding high-frequency-expansion
  (see section 3.2 and Appendix A  of the the review \cite{floquet_review})
for the Kick operator $K_{eff}(t_0)$ that depends on the Floquet gauge $t_0$ ($t_0$ is the initial time chosen to define the 
stroboscopic one-period dynamics on $[t_0,t_0+T]$) and that describes the micromotion withing the driving period $T=\frac{2 \pi}{\Omega}$. While this micromotion withing the driving period  is very important is many Floquet applications and requires
the computation of the Kick operator $K_{eff}(t_0)$, we will not be interested in this micromotion within our present
renormalization framework in the time domain. Indeed, our goal will be to eliminate iteratively the higher frequency $\Omega$ 
 (i.e. the smaller period $T$) in order to derive the effective dynamics on time scales much larger than $T$,
so that we will disregard all dynamical properties associated to time scales shorter than $T$.

Let us finish with a small technical remark : in the following sections, 
it will be convenient to consider also the case of 'negative frequency' $\Omega<0$.
In the above formula, if $\Omega>0$ is changed into 
\begin{eqnarray}
 {\tilde \Omega}=-\Omega <0
\label{negom}
\end{eqnarray}
then the Fourier components of Eq \ref{hfourier} are changed into
\begin{eqnarray}
{\tilde  {\cal H}}_l = {\cal H}_{(-l)} 
\label{hnegom}
\end{eqnarray}
As a consequence, Eq \ref{heff0} corresponding to the Fourier mode $l=0$
is unchanged, while Eq. \ref{heff1} is also unchanged because both the denominator $\Omega$ and the commutator $ [ {\cal H}_l, {\cal H}_{-l} ]$
present a change of sign. In conclusion, the formula of Eq. \ref{heff} for the
 effective Floquet Hamiltonian can be applied both for positive or negative frequency $\Omega$.

\subsection{ Application to the local degree of freedom having the highest eigenfrequency }

The Vosk-Altman Strong Disorder Renormalization \cite{vosk_dyn1,vosk_dyn2} for the unitary dynamics 
within the Many-Body-Localized phase of interacting random systems can be reformulated as follows :
when the distribution of eigenfrequencies of the local degrees of freedom is sufficiently broad,
 the local degree of freedom having the biggest eigenfrequency $\Omega$
can be considered as a high-frequency-Floquet-driving for the neighboring slower degrees of freedom,
so that one can apply the general Floquet theory recalled above.

In disordered models, each random coupling $K_i$  is generically associated to a local degree of freedom
with only two energy-levels $e_1$ and $e_2$ associated to two corresponding projectors $P_1$ and $P_2$
\begin{eqnarray}
H^{TwoLevels}_{K_i} = e_1 P_1+e_2 P_2
\label{hki}
\end{eqnarray}
 so that the difference between these two energies defines the eigenfrequency
of the local degree of freedom associated to the random coupling $K_i$
\begin{eqnarray}
\Omega_{K_i} = e_2-e_1
\label{hki}
\end{eqnarray}
It turns out that in one example of application, we will also encounter a special case
where the local Hamiltonian associated to some random coupling has instead
three equally-spaced levels $(e_1,e_2,e_3)$ 
so that the eigenfrequency corresponds to $\Omega=e_3-e_2=e_2-e_1$.
To consider these two cases within the same framework,
it is thus convenient to parametrize the Hamiltonian $H_0$
of the local degree of freedom having the biggest eigenfrequency $\Omega$
 via the following spectral decomposition into $M$ levels $m=1,..,M$ of energies $e_m=e_0+m \Omega$
where $e_0$ is some constant
\begin{eqnarray}
H_0= \sum_{m=1}^M (e_0 + m \Omega) P_m
\label{h0proj}
\end{eqnarray}
where the corresponding orthogonal projectors projectors $P_m$ satisfy
\begin{eqnarray}
P_{m} P_{m'} = P_m \delta_{m,m'}
\label{orthogproj}
\end{eqnarray}

In the interaction picture (see your favorite quantum mechanics textbook), the rest of the Hamiltonian
\begin{eqnarray}
V\equiv H-H_0
\label{vreste}
\end{eqnarray}
becomes the time-periodic Hamiltonian
\begin{eqnarray}
V^{inter} (t) && \equiv e^{i H_0 t} V e^{-i H_0 t} = \left(  \sum_{m=1}^M e^{i  (e_0 + m \Omega) t}  P_m \right) V 
\left(  \sum_{m'=1}^M e^{- i  (e_0 + m' \Omega) t }  P_{m'} \right)
\nonumber \\
&& =   \sum_{m=1}^M \sum_{m'=1}^M e^{i  (m-m')  \Omega t }  P_m V   P_{m'} 
\label{vintert}
\end{eqnarray}
so that the Fourier decomposition of Eq \ref{hfourier} contains only a small number of Fourier modes $l=m-m' = -(M-1),..,+M-1$
\begin{eqnarray}
V^{inter}(t) = \sum_{l =-(M-1)}^{M-1} {\cal V}_l e^{i l \Omega t}
\label{vfourier}
\end{eqnarray}
with the Fourier coefficients
\begin{eqnarray}
{\cal V}_l  && =   \sum_{m=1}^M \sum_{m'=1}^M  P_m V   P_{m'} \delta_{l,m-m'}
\label{vlfourier}
\end{eqnarray}
In particular, the coefficient for $l=0$ contains the operator $V$ sandwiched between two identical projectors $m=m'$
\begin{eqnarray}
{\cal V}_{l=0}  && =   \sum_{m=1}^M   P_m V   P_{m} = P_1 V   P_1+ ... + P_M V   P_{M} 
\label{fourierzero}
\end{eqnarray}
while the coefficients for positive $l=1,..,M-1$ contains the operator $V$ sandwiched between two different projectors $m'=m-l<m$
\begin{eqnarray}
{\cal V}_{l>0}  && =   \sum_{m=l+1}^M   P_m V   P_{m-l} = P_{l+1} V   P_{1}  + .. P_M V   P_{M-l}
\label{vlposi}
\end{eqnarray}
and the coefficient for $(-l) $ corresponds to $m=m'-l<m'$
\begin{eqnarray}
{\cal V}_{-l<0}  && =  \sum_{m'=l+1}^M  P_{m'-l} V   P_{m'} =P_{1} V   P_{l+1}  + .. P_{M-l} V   P_{M} =   {\cal V}_{l}^{\dagger} 
\label{vlnega}
\end{eqnarray}

The commutators needed in the Floquet effective Hamiltonian (Eq. \ref{heff1}) read for $l=1,..,M-1$
 using the properties of the orthogonal projectors $P_m$ (Eq. \ref{orthogproj})
\begin{eqnarray}
 [ {\cal V}_l, {\cal V}_{-l} ] && =\sum_{m=l}^{M-1}  \sum_{m'=l}^{M-1}   [  P_m V   P_{m-l}  ,   P_{m'-l} V   P_{m'}  ] 
\nonumber \\
&& = \sum_{m=l}^{M-1}  \sum_{m'=l}^{M-1}   (  P_m V   P_{m-l}    P_{m'-l} V   P_{m'}  -     P_{m'-l} V   P_{m'} P_m V   P_{m-l} )
\nonumber \\
&& = \sum_{m=l}^{M-1}     (  P_m V   P_{m-l}    V   P_{m}  -     P_{m-l} V   P_m V   P_{m-l} )
\label{vlcomm}
\end{eqnarray}

The final result for the Floquet effective Hamiltonian $V_{eff}$ of Eqs \ref{heff} \ref{heff0} \ref{heff1} thus reads using Eqs \ref{fourierzero} and \ref{vlcomm}
\begin{eqnarray}
V_{eff} && = V_{eff}^{(0)}+  V_{eff}^{(1)} + O\left(\frac{1}{\Omega^2} \right)
\nonumber \\
&& = {\cal V}_0 + \frac{1}{\Omega} \sum_{l =1}^{M-1} \frac{1}{l} [ {\cal V}_l, {\cal V}_{-l} ]+ O\left(\frac{1}{\Omega^2} \right)
\nonumber \\
&& =  \sum_{m=1}^M   P_m V   P_{m}   + \sum_{l =1}^{M-1} \frac{1}{l \Omega}
 \sum_{m=l+1}^M     (  P_m V   P_{m-l}    V   P_{m}  -     P_{m-l} V   P_m V   P_{m-l} )
+ O\left(\frac{1}{\Omega^2} \right)
\label{veffproj}
\end{eqnarray}
This equation is the central recipe that allows
to obtain the renormalization rules for the dynamics of the slower degrees of freedom in various models.
Before we turn to specific examples, it is useful to discuss first the general physical meaning of this effective Floquet Hamiltonian
and the links with other approaches.

\subsection { Emergent Local Integrals of Motion (LIOMs)   }

\label{sec_emerg}

The most important property of the effective Floquet Hamiltonian of Eq. \ref{veffproj}
is that it commutes with the set of the orthogonal projectors $P_m$ (Eq. \ref{orthogproj})
\begin{eqnarray}
[V_{eff} , P_m ]&& = 0
\label{cveffproj}
\end{eqnarray}
In the present framework, these projectors $P_m$ onto the energy levels of the Hamiltonian $H_0$
(Eq.  \ref{h0proj}) describing the degree of freedom having the biggest high-frequency
are thus the emergent Local Integrals of Motion (LIOMs). mentioned in the Introduction.

The physical meaning is that the neighboring slow degrees of freedom described by the Floquet effective Hamiltonian $V_{eff}$
are not able to exchange energy with this high-frequency degree of freedom described by $H_0$.
As a consequence, the energy levels of $H_0$ as described by the projectors $P_m$ are effectively conserved by the dynamics.
The first term of Eq. \ref{veffproj} associated to the Fourier mode $l=0$
corresponds simply to the projection of $V$ onto the energy levels of $H_0$.
The second term of Eq. \ref{veffproj} associated to the Fourier modes $l=1,..,M$
contains virtual processes between the different energy levels separated by the energy difference $(l \Omega)$
that appear in the denominator : these new renormalized contributions will thus be small for large frequency $\Omega$.

We should now discuss the dimensionality of the projectors $P_m$ :
 if the projector $P_m$ is one-dimensional, then this conserved degree of freedom simply disappears
in the effective description of the slower degrees of freedom once its effects have been taken into account in $V_{eff}$
(see the example in section \ref{sec_h});
 if the projector $P_m$ is instead two-dimensional or higher, then one needs to introduce a new renormalized degree of freedom
to label the degenerate levels associated to the same projector $P_m$ (see the example in section \ref{sec_j}).

\subsection { Physical meaning of the decomposition into fast and slow modes   }

In conclusion, the Strong Disorder RG approach for the dynamics can be summarized as follows.
At some given time $t$, the degrees of freedom are separated into two groups with respect to $\Omega_t = \frac{1}{t} $ :

(i) the degrees of freedom that would have had higher eigenfrequencies $\vert \Omega \vert > \Omega_t$
have been converted into LIOMs via the projectors $P_m$ : in some sense, they have converged towards 
their asymptotic state described by the diagonal ensemble $m=m'$ of their local Hamiltonian $H_0$,
while the off-diagonal contributions $m \ne m'$ have been time-averaged-out.

(ii) the remaining degrees of freedom that are characterized by renormalized eigenfrequencies $\vert \Omega \vert < \Omega_t$
have not yet converged towards their asymptotic state, since they have not had enough time to oscillate with their eigenfrequency.

At a qualitative level, this dichotomy is reminiscent of the
 Strong-Disorder-RG procedures for thermally-activated-classical dynamics
(see the review \cite{strong_review})
concerning either random classical spin chains \cite{sinaiprl,rfim},
or various random walks in random media \cite{sinaiprl,sinaipre,sinaireaction,golosov,landscape,energysinai,trapdirected,trapsymmetric,trapreponse}.
In these studies, the dynamics is also decomposed into two parts according to the time needed to overcome Arrhenius dynamical barriers :
at some given time $t$, the fast degrees of freedom have already converged towards their local equilibrium,
while the slow degrees of freedom are still completely out-of-equilibrium.

\subsection { Equivalence with the RSRG-X procedure to construct the eigenstates    }

The effective Floquet Hamiltonian of Eq. \ref{veffproj} can be decomposed into its contributions on the various energy-levels selected by the projectors $P_m$
representing the LIOMs
\begin{eqnarray}
V_{eff}  
&& =  \sum_{m=1}^M   P_m V_{eff}   P_{m}   
+ O\left(\frac{1}{\Omega^2} \right)
\label{veffsumlevel}
\end{eqnarray}
Using the energies of Eq. \ref{h0proj}, the contribution in each energy-level can be rewritten using the energies of Eq. \ref{h0proj}
\begin{eqnarray}
P_m V_{eff} P_m && 
 =    P_m V   P_{m}   + P_m V \left(  \sum_{m' \ne m} \frac{P_{m'}}{ E_m-E_{m'} } \right) V P_m
\nonumber \\
&& 
 =    P_m V   P_{m}   + P_m V (1-P_m)  \frac{1}{ E_m- H_0 } (1-P_m) V P_m
\label{pereachlevel}
\end{eqnarray}
One recognizes the two first orders of the standard perturbation theory for the energy-levels of $H_0$ :
the first order in the perturbation $V$ is simply the projection $ P_m V   P_{m}   $ on the energy-level $E_m$,
while the second order in the perturbation $V$ involves the virtual processes towards the other energy levels $m' \ne m$,
with the usual energy differences $ (E_m-E_{m'} ) $  in the denominators.

With this reformulation, the equivalence with the the RSRG-X procedure introduced in \cite{rsrgx} to construct the whole set of excited eigenstates 
is now obvious : at each step, one chooses the local degree of freedom described by the local Hamiltonian $H_0$ with the biggest gap, i.e. the
biggest energy differences. The energy-levels of $H_0$ are then considered as LIOMs, and the effective Hamiltonian for the 
remaining degrees of freedom is computed via the second order perturbation formula in each level via Eq. \ref{pereachlevel}.
Here the link with
the Fisher Strong Disorder Real-Space RG \cite{fisher_AF,fisher,fisherreview} to construct the ground states  (see the review \cite{strong_review}) is also 
extremely clear : to construct the ground-state, one projects only onto the lowest-energy-level of $H_0$ at each step
with the same formula of Eq. \ref{pereachlevel}, while in the RSRG-X, one considers in parallel all the energy-levels of $H_0$ with the same formula of Eq. \ref{pereachlevel}.

In summary, the RSRG-t to construct the effective dynamics and the RSRG-X to construct the set of eigenstates
are equivalent in practice to derive the renormalization rules via Eqs \ref{veffproj} and \ref{pereachlevel},
but the two formulations are nevertheless both useful since they provide complementary points of view.

\subsection{ Application to the Random transverse field XXZ chain  }

As an example of application of the general procedure described above, let us consider
the XXZ chain involving Pauli matrices with random transverse fields $h_j$,
with couplings $J_j^x=J_j^y=J_j$ (hopping in the Jordan-Wigner fermionic formulation) 
and $J_j^z = \Delta_j$ (interaction in the Jordan-Wigner fermionic formulation) 
\begin{eqnarray}
H && = \sum_{j=1}^N \left[ h_j \sigma_j^z +\Delta_j \sigma_j^z \sigma_{j+1}^z
 +  J_j ( \sigma_j^x \sigma_{j+1}^x+\sigma_j^y \sigma_{j+1}^y ) \right] 
\nonumber \\
&&  = \sum_{j=1}^N \left[ h_j \sigma_j^z + \Delta_j \sigma_j^z \sigma_{j+1}^z
 + 2 J_j ( \sigma_j^+ \sigma_{j+1}^-+\sigma_j^- \sigma_{j+1}^+ ) \right] 
\label{heisen}
\end{eqnarray}
The standard model considered in most numerical studies \cite{kjall,alet,alet_dyn,luitz_tail,badarson_signa,auerbach,znidaric_dephasing,prelo_dyn,znidaric_lindblad,
luitz_bimodal,garcia,luitz_anomalous,luitz_operator,luitz_information} corresponds to homogeneous couplings $J_j=J=1$ and interaction $\Delta_j=\Delta=1$,
while the transverse fields $h_j$ are random variables drawn uniformly on $[-h_0,h_0]$.
  Then the Many-Body-Localized phase has been found to exist for sufficiently large disorder $h_0 > h_c \simeq 3.7$ via various criteria.

Here we wish to analyze the dynamics in the Many-Body-Localized phase far from the critical point $h_0 \gg h_c$.
It is then clear that the first renormalization steps will only concern the large transverse fields $h_j$.
The corresponding RG rules are analyzed in section \ref{sec_h} below
and yield that the renormalized couplings $J_j$ and interactions $\Delta_j$ become different $J_j \ne \Delta_j$
and depend on the position $j$ along the chain,
even if one starts from the usual homogeneous case $J=\Delta=1$.
Then the question arises as to what should be done if after these many field-decimations, the biggest 
remaining eigenfrequency is associated to some coupling $J_j$ or to some interaction $\Delta_j$.
When the biggest eigenfrequency is associated to some coupling $J_j$, the renormalization rules
are analyzed in section \ref{sec_j}.
When the biggest eigenfrequency is associated to some interaction $\Delta_j$, the renormalization rules
can also be written, but they generate new terms with respect to the Hamiltonian of Eq. \ref{heisen}. 
The full analysis of the renormalization flow that would include all these new-generated terms 
clearly goes beyond the scope of the present work.
To maintain a closed RG flow within the Hamiltonians of the form of Eq. \ref{heisen},
we will thus assume that the biggest eigenfrequency $\Omega$ is associated either to a field $h_j$ or to a coupling $J_j$,
but never to an interaction $\Delta_j$.
Since the case without interaction $\Delta_j=0$ corresponds to a random free-fermion Hamiltonian that is always
in the localized phase, the following RG procedure 
 will allow to analyze the stability of this Localized phase in the presence of small interactions $\Delta_j$.

In summary, the renormalization procedure that we consider in the remainder of this paper can be summarized as follows.
At some given stage of the RG procedure, the effective Floquet Hamiltonian has the form of Eq. \ref{heisen} up to constant terms involving only the previously generated LIOMs :

(a) Each remaining transverse field $h_j$ defines the natural eigenfrequency 
\begin{eqnarray}
\Omega_{(h_j)} = 2  h_j 
\label{omegaa}
\end{eqnarray}
that would govern the precessing of the single spin $\sigma_j$ if it were only submitted to $H_0= h_j \sigma_j^z  $.

(b) Each remaining coupling $J_j$  defines the natural eigenfrequency
\begin{eqnarray}
\Omega_{(J_j)} = 2  J_j 
\label{omegac}
\end{eqnarray}
that would appear if the two spins $(\sigma_j , \sigma_{j+1} )$ were only submitted to $H_0= 2 J_j ( \sigma_j^+ \sigma_{j+1}^-+\sigma_j^- \sigma_{j+1}^+ )$.

One chooses the biggest eigenfrequency in absolute value $\vert \Omega \vert $ remaining the chain :
if this biggest frequency corresponds to a transverse-field (case (a) of Eq. \ref{omegaa}),
one applies the RG rules derived in section \ref{sec_h};
if this biggest frequency corresponds to a coupling (case (b) of Eq. \ref{omegac}),
one applies the RG rules derived in section \ref{sec_j}.

As a final remark, let us mention that the two RSRG-t papers by Vosk and Altman 
concerns respectively the random XXZ without random fields (i.e. the special case $h_j=0$ in Eq. \ref{heisen})  \cite{vosk_dyn1}
and the weakly-interacting transverse Ising model (i.e. $J_j^{yy}=0$ and $\Delta_j \ll J_j^{xx} $) \cite{vosk_dyn2}.

\section { RG rules when the highest eigenfrequency is $\Omega=2  h_n  $ }

\label{sec_h}

In this section, we consider the case where the highest eigenfrequency remaining in the chain is associated to a transverse field (Eq. \ref{omegaa})
\begin{eqnarray}
\Omega = 2  h_n 
\label{omegaan}
\end{eqnarray}

\subsection{ Properties of $H_0$ }

The corresponding Hamiltonian of Eq. \ref{h0proj}
\begin{eqnarray}
H_0= h_n \sigma_n^z  
\label{h0projhn}
\end{eqnarray}
is associated to the single spin $\sigma_n$, so that the Hilbert space is two-dimensional.
In the notations of Eq. \ref{h0proj},  $H_0$ has $M=2$ non-degenerate levels $(\pm  h_n )$
associated to the projectors
\begin{eqnarray}
P_1 && =  \frac{1- \sigma_n^z}{2} = \vert \sigma_n^z= -><\sigma_n^z=- \vert
\nonumber \\
P_2 &&  = \frac{1+ \sigma_n^z}{2}= \vert \sigma_n^z= +><\sigma_n^z =+ \vert
\label{hnproj}
\end{eqnarray}

\subsection{Effective Floquet Hamiltonian for the rest $V=H-H_0$  }

The rest of the chain $V=H-H_0$ 
has for effective Floquet Hamiltonian (Eq. \ref{veffproj})
\begin{eqnarray}
V_{eff}  
&& =  \sum_{m=1}^2   P_m V   P_{m}   + \sum_{l =1}^{1} \frac{1}{l \Omega}
 \sum_{m=2}^2     (  P_m V   P_{m-l}    V   P_{m}  -     P_{m-l} V   P_m V   P_{m-l} )
+ O\left(\frac{1}{\Omega^2} \right)
\nonumber \\
&& =    P_1 V   P_1 + P_2 V   P_2  +  \frac{1}{ \Omega}
      (  P_2 V   P_{1}    V   P_{2}  -     P_{1} V   P_2 V   P_{1} )
+ O\left(\frac{1}{\Omega^2} \right)
\label{veffproja}
\end{eqnarray}

It is thus convenient to decompose $V$ into the various contributions involving the spin $\sigma_n$
\begin{eqnarray}
V\equiv H-H_0 ={\cal B}_n^0 + {\cal B}_n^z \sigma_n^z 
 +{\cal B}_n^+ \sigma_n^+  
+ {\cal B}_n^- \sigma_n^-  
\label{vresteb}
\end{eqnarray}
with the operators depending on the neighboring spins
\begin{eqnarray}
{\cal B}_n^z && = \Delta_{n-1} \sigma_{n-1}^z + \Delta_n \sigma_{n+1}^z 
\nonumber \\
{\cal B}_n^+ && =  2J_{n-1} \sigma_{n-1}^- + 2J_n  \sigma_{n+1}^- 
\nonumber \\
{\cal B}_n^- && =   2J_{n-1} \sigma_{n-1}^+ + 2J_n  \sigma_{n+1}^+
\label{calB}
\end{eqnarray}
while $ {\cal B}_n^0 $ contains all the terms of Eq \ref{heisen} that do not involve the spin $\sigma_n$.

Using the two projectors of Eq. \ref{hnproj}, Eq. \ref{veffproja} becomes
\begin{eqnarray}
V_{eff}  ={\cal B}_n^0 + {\cal B}_n^z (P_2-P_1) 
+ \frac{1}{ \Omega}     (  P_2 {\cal B}_n^+ {\cal B}_n^-   P_{2}  -     P_{1} {\cal B}_n^- {\cal B}_n^+   P_{1} )  
\label{veffhn}
\end{eqnarray}
so we need to evaluate
\begin{eqnarray}
{\cal B}_n^+ {\cal B}_n^-  && =  (2J_{n-1} \sigma_{n-1}^- + 2J_n  \sigma_{n+1}^- ) (2J_{n-1} \sigma_{n-1}^+ + 2J_n  \sigma_{n+1}^+)
\nonumber \\
&& = 2 J_{n-1}^2 (1-\sigma_{n-1}^z) + 2 J_n^2 (1-\sigma_{n+1}^z) + 4 J_{n-1} J_n (\sigma_{n-1}^-\sigma_{n+1}^+
 +\sigma_{n-1}^+ \sigma_{n+1}^- )
\nonumber \\
{\cal B}_n^- {\cal B}_n^+&& =   (2J_{n-1} \sigma_{n-1}^+ + 2J_n  \sigma_{n+1}^+)(2J_{n-1} \sigma_{n-1}^- + 2J_n  \sigma_{n+1}^- )
\nonumber \\
&& = 2 J_{n-1}^2 (1+\sigma_{n-1}^z) + 2 J_n^2 (1+\sigma_{n+1}^z) + 4 J_{n-1} J_n (\sigma_{n-1}^-\sigma_{n+1}^+
 +\sigma_{n-1}^+ \sigma_{n+1}^- )
\label{calBb}
\end{eqnarray}

As explained in section \ref{sec_emerg}, the projectors $P_1$ and $P_2$ are the emergent Local Integrals of Motion (LIOMs) 
produced by this renormalization step.
Putting everything together, one finally
 obtains that the effective Floquet Hamiltonian for the remaining degrees of freedom reads
\begin{eqnarray}
V_{eff}  && = {\cal B}_n^0 +  (  P_2 -P_1) \left(  \frac{ J_{n-1}^2 +J_n^2   }{  h_n } \right) 
\nonumber \\
&& + \left(  (P_2-P_1 )  \Delta_{n-1}- \frac{ J_{n-1}^2}{   h_n  }    \right)    \sigma_{n-1}^z       
+  \left(  (P_2-P_1 ) \Delta_n- \frac{ J_n^2}{   h_n }    \right)     \sigma_{n+1}^z      
\nonumber \\
&& +   2   (  P_2     -     P_{1}    )  
 \frac{ J_{n-1} J_n}{   h_n  } (\sigma_{n-1}^-\sigma_{n+1}^+ +\sigma_{n-1}^+ \sigma_{n+1}^- )
\label{vhnfinal}
\end{eqnarray}

\subsection{ Renormalization rules  }

The effective Floquet Hamiltonian of Eq. \ref{vhnfinal} means that the form of XXZ Hamiltonian of Eq. \ref{heisen} is preserved,
up to constant terms involving only the projectors,
with the following renormalized parameters :

(i) the transverse fields on the two neighboring spins $\sigma_{n-1}^z$ and $\sigma_{n+1}^z $ are renormalized into
\begin{eqnarray}
h_{n-1}^R && = h_{n-1}+ (P_2-P_1 )  \Delta_{n-1} -   \frac{J_{n-1}^2}{  h_n } 
\nonumber \\
h_{n+1}^R && = h_{n+1} + (P_2-P_1 ) \Delta_n - \frac{ J_{n}^2}{ h_n } 
\label{hra}
\end{eqnarray}

(ii)  the new renormalized hopping between these two spins 
\begin{eqnarray}
J^R_{n-1,n+1} =  (  P_2     -     P_{1}    )  \frac{ J_{n-1} J_n}{  h_n } 
\label{jra}
\end{eqnarray}
is expected to be small since the transverse-field $h_n$ has been chosen as the biggest coupling remaining in the chain.

(iii)  no interaction term is generated between these two spins
\begin{eqnarray}
\Delta^R_{n-1,n+1} = 0
\label{dra}
\end{eqnarray}

This largest-transverse-field decimation step has thus for effects to generate small renormalized hopping $J^R$ and vanishing interaction $\Delta^R$.
For large initial disorder where this renormalization step is chosen most often, it is clear that the renormalization flow is towards
the Many-Body-Localized phase. Nevertheless, it is interesting in the next section to consider the effects
of the rare decimations concerning the couplings.

\section { RG rules when the highest eigenfrequency is $\Omega=2  J_n  $ }

\label{sec_j}

In this section, we consider the case where the highest eigenfrequency remaining in the chain is associated to 
the coupling $J_n$ (Eq. \ref{omegac})
\begin{eqnarray}
\Omega = 2  J_n 
\label{omegabn}
\end{eqnarray}

\subsection{ Properties of $H_0$ }

The corresponding Hamiltonian of Eq. \ref{h0proj}
\begin{eqnarray}
H_0= 2 J_n  (\sigma_n^+  \sigma_{n+1}^- + \sigma_{n}^-\sigma_{n+1}^+)
\label{h0projjn}
\end{eqnarray}
is associated to the pair of spins $(\sigma_n,\sigma_{n+1})$, so that the Hilbert space is four-dimensional.
In the notations of Eq. \ref{h0proj},  $H_0$ has $M=3$ levels $(-2 J_n, 0, +2 J_n)$.
The middle level of zero-energy is degenerate twice
\begin{eqnarray}
P_{2} && = \vert ++><++ \vert + \vert --><-- \vert = \frac{1+\sigma_n^z \sigma_{n+1}^z}{2}
\label{proj2}
\end{eqnarray}
while the two other energy levels are non-degenerate with the projectors
\begin{eqnarray}
P_{1}&&=   \vert \psi_{1}>  <   \psi_{1}\vert
=  \frac{1-\sigma_n^z \sigma_{n+1}^z}{4} - \frac{\sigma_n^+  \sigma_{n+1}^- + \sigma_{n}^-\sigma_{n+1}^+}{2}
\nonumber \\
P_{3}&&= \vert \psi_{3}>  <   \psi_{3}\vert
= \frac{1-\sigma_n^z \sigma_{n+1}^z}{4} + \frac{\sigma_n^+  \sigma_{n+1}^- + \sigma_{n}^-\sigma_{n+1}^+}{2}
\label{proj13}
\end{eqnarray}
associated to the kets
\begin{eqnarray}
\vert \psi_{1}>  &&=  \frac{\vert +->-\vert -+>} {\sqrt 2}
\nonumber \\
\vert \psi_{3}>   &&= \frac{\vert +->+ \vert -+>} {\sqrt 2}
\label{ket}
\end{eqnarray}

Since $P_2$ is degenerate, one needs to introduce a renormalized spin $\sigma_R$ to take into account 
this degree of freedom within this energy level
\begin{eqnarray}
\vert \sigma_R^z =+ > && = \vert ++>
\nonumber \\
\vert \sigma_R^z =- > && =  \vert -->
\label{projsigmar}
\end{eqnarray}

\subsection{Effective Floquet Hamiltonian for the rest $V=H-H_0$  }

The rest of the chain $V=H-H_0$ 
has for effective Floquet Hamiltonian (Eq. \ref{veffproj})
\begin{eqnarray}
V_{eff}  
&& =  \sum_{m=1}^3   P_m V   P_{m}   + \sum_{l =1}^{2} \frac{1}{l \Omega}
 \sum_{m=l+1}^3     (  P_m V   P_{m-l}    V   P_{m}  -     P_{m-l} V   P_m V   P_{m-l} )
+ O\left(\frac{1}{\Omega^2} \right)
\nonumber \\
&& =  (   P_1 V   P_1 + P_2 V   P_2 + P_3 V   P_3 )
+ \frac{1}{ \Omega}
\left(        P_2 V   P_{1}    V   P_{2}  -     P_{1} V   P_2 V   P_{1} 
+         P_3 V   P_{2}    V   P_{3}  -     P_{2} V   P_3 V   P_{2} 
\right)
\nonumber \\
&& +  \frac{1}{2 \Omega}    (  P_3 V   P_{1}    V   P_{3}  -     P_{1} V   P_3 V   P_{1} )
+ O\left(\frac{1}{\Omega^2} \right)
\label{veffprojc}
\end{eqnarray}

Here we need to decompose $V$ into the various contributions involving the spins $\sigma_n$ and $\sigma_{n+1}$
\begin{eqnarray}
V\equiv H-H_0 && ={\cal B}_n^0 + \Delta_n \sigma_n^z  \sigma_{n+1}^z
\nonumber \\
&&  +{\cal B}_n^z \sigma_n^z 
+{\cal B}_{n+1}^z \sigma_{n+1}^z   +   {\cal B}_n^- \sigma_n^- +{\cal B}_n^+ \sigma_n^+
+{\cal B}_{n+1}^-  \sigma_{n+1}^- +{\cal B}_{n+1}^+ \sigma_{n+1}^+  
\label{vhnc}
\end{eqnarray}
with the operators depending on the other spins
\begin{eqnarray}
{\cal B}_n^z && = h_n + \Delta_{n-1} \sigma_{n-1}^z 
\nonumber \\
{\cal B}_{n+1}^z && =
 h_{n+1} + \Delta_{n+1} \sigma_{n+2}^z 
\nonumber \\
{\cal B}_n^- && =   2J_{n-1} \sigma_{n-1}^+
\nonumber \\
{\cal B}_n^+ && = 2J_{n-1} \sigma_{n-1}^-
\nonumber \\
{\cal B}_{n+1}^- && =  2 J_{n+1} \sigma_{n+2}^+ 
\nonumber \\
{\cal B}_{n+1}^+ && = 2 J_{n+1} \sigma_{n+2}^- 
\label{calBc}
\end{eqnarray}
while  $ {\cal B}_n^0 $ contains all the terms of Eq. \ref{heisen} that do not involve the spins $\sigma_n$ and $\sigma_{n+1}$.

Since the term involving the two spins can be rewritten in terms of the projectors
\begin{eqnarray}
 \sigma_n^z \sigma_{n+1}^z = P_2-P_{1} -P _{3}  
\label{v0jn}
\end{eqnarray}
we only need to compute the matrix element of the terms of Eq. \ref{vhnc} involving a single spin operator
\begin{eqnarray}
V_{single} && ={\cal B}_n^z \sigma_n^z 
+{\cal B}_{n+1}^z \sigma_{n+1}^z  
 +   {\cal B}_n^- \sigma_n^- 
+{\cal B}_n^+ \sigma_n^+
+{\cal B}_{n+1}^-  \sigma_{n+1}^- 
+{\cal B}_{n+1}^+ \sigma_{n+1}^+  
\label{vhn}
\end{eqnarray}

Using its action on the four vector basis
\begin{eqnarray}
V_{single} \vert \psi_{1}>  &&=
({\cal B}_n^z   - {\cal B}_{n+1}^z )  \vert \psi_{3}>
 +    \left( \frac{  {\cal B}_n^- -{\cal B}_{n+1}^-   } {\sqrt 2} \right) \vert -->
+ \left( \frac{{\cal B}_{n+1}^+-{\cal B}_n^+ } {\sqrt 2} \right)  \vert ++>
\nonumber \\
V_{single} \vert \psi_{3}>   &&= ({\cal B}_n^z - {\cal B}_{n+1}^z ) \vert \psi_{1}>
 +   \left( \frac{ {\cal B}_n^-  +{\cal B}_{n+1}^-     } {\sqrt 2} \right) \vert -->
+\left( \frac{{\cal B}_{n+1}^+ +{\cal B}_n^+ } {\sqrt 2} \right) \vert ++>
\nonumber \\
V_{single} \vert ++ >  &&= ( {\cal B}_n^z  +{\cal B}_{n+1}^z ) \vert ++ > 
+ \left( \frac{{\cal B}_{n+1}^- -{\cal B}_n^- } {\sqrt 2} \right)  \vert \psi_1>
+\left( \frac{{\cal B}_{n+1}^- +{\cal B}_n^-  } {\sqrt 2} \right) \vert \psi_3>
\nonumber \\
V_{single} \vert -- >   &&= - ( {\cal B}_n^z  + {\cal B}_{n+1}^z  ) \vert -- >
 +    \left( \frac{  {\cal B}_n^+ -{\cal B}_{n+1}^+  } {\sqrt 2} \right) \vert \psi_1>
+   \left( \frac{ {\cal B}_n^+  +{\cal B}_{n+1}^+    } {\sqrt 2} \right) \vert \psi_3 >
\label{vsinglemat}
\end{eqnarray}
one finally obtains with $\Omega=2  J_n  $ and the renormalized spin of Eq. \ref{projsigmar}
\begin{eqnarray}
V_{eff}  && = {\cal B}_n^0 + \Delta_n (P_2-P_{1} -P _{3}   )  + \frac{ (J_{n-1}^2+J_{n+1}^2)}{   J_n  } (P_3-P_1)
+  \frac{(h_n-h_{n+1})^2 + \Delta_{n-1}^2+\Delta_{n+1}^2}{4  J_n }   (  P_3 -   P_{1}    )
\nonumber \\
&& 
+ ( h_n + \Delta_{n-1} \sigma_{n-1}^z   + h_{n+1} + \Delta_{n+1} \sigma_{n+2}^z  ) \sigma_R^z
\nonumber \\
&& 
+2 \frac{ J_{n-1} J_{n+1}  }{   J_n  } (P_3+P_1- P_2)  ( \sigma_{n-1}^+ \sigma_{n+2}^-+\sigma_{n-1}^- \sigma_{n+2}^+  ) 
\nonumber \\
&& +  \frac{ (h_n-h_{n+1}) ( \Delta_{n-1} \sigma_{n-1}^z - \Delta_{n+1} \sigma_{n+2}^z ) - \Delta_{n-1} \Delta_{n+1} \sigma_{n-1}^z\sigma_{n+2}^z}{ 2  J_n  }   (  P_3 -   P_{1}    )
+ O\left(\frac{1}{\Omega^2} \right)
\label{veffprojcbis}
\end{eqnarray}

Since the renormalized spin of Eq. \ref{projsigmar} only appears via $\sigma_R^z$, 
it is actually also an emergent Local Integral of Motion (Liom)
\begin{eqnarray}
[V_{eff}  , \sigma_R^z ] =0
\label{veffprojcom}
\end{eqnarray}
It can be thus eliminated from the effective Floquet Hamiltonian $V_{eff}$ via the use of the projectors
\begin{eqnarray}
P_{2+} && = \vert ++><++ \vert 
\nonumber \\
P_{2-} && = \vert --><-- \vert 
\label{proj2pm}
\end{eqnarray}
while it is useful to keep the notation
\begin{eqnarray}
P_2=P_{2+} + P_{2-} 
\label{proj2pmsum}
\end{eqnarray}

In summary, the two spins $\sigma_n$ and $\sigma_{n+1}$ are completely taken into account via the four one-dimensional projectors
$(P_1,P_{2+},P_{2-},P_3)$ that represent the Lioms generated by this renormalization step.
The effective Floquet Hamiltonian for the remaining spins of the chain reads
\begin{eqnarray}
V_{eff}  && = {\cal B}_n^0 + \Delta_n (P_2-P_{1} -P _{3}   )  + \frac{ (J_{n-1}^2+J_{n+1}^2)}{   J_n } (P_3-P_1)
+  \frac{(h_n-h_{n+1})^2 + \Delta_{n-1}^2+\Delta_{n+1}^2}{4  J_n  }   (  P_3 -   P_{1}    )
\nonumber \\
&& 
+ ( h_n + \Delta_{n-1} \sigma_{n-1}^z   + h_{n+1} + \Delta_{n+1} \sigma_{n+2}^z  ) (P_{2+} -P_{2-}  )
\nonumber \\
&& 
+2 \frac{ J_{n-1} J_{n+1}  }{   J_n } (P_3+P_1- P_2)  ( \sigma_{n-1}^+ \sigma_{n+2}^-+\sigma_{n-1}^- \sigma_{n+2}^+  ) 
\nonumber \\
&& +  \frac{ (h_n-h_{n+1}) ( \Delta_{n-1} \sigma_{n-1}^z - \Delta_{n+1} \sigma_{n+2}^z ) - \Delta_{n-1} \Delta_{n+1} \sigma_{n-1}^z\sigma_{n+2}^z}{ 2 J_n  }   (  P_3 -   P_{1}    )
+ O\left(\frac{1}{\Omega^2} \right)
\label{vefffinal}
\end{eqnarray}

\subsection{ Renormalization rules  }

The effective Floquet Hamiltonian of Eq. \ref{vhnfinal} means that the form of XXZ Hamiltonian of Eq. \ref{heisen} is preserved,
up to constant terms involving only the projectors,
with the following renormalized parameters :

(i) the transverse fields on the two neighboring spins $\sigma_{n-1}^z$ and $\sigma_{n+2}^z $ are renormalized into
\begin{eqnarray}
h_{n-1}^R&&= h_{n-1} +  \Delta_{n-1}  (P_{2+} -P_{2-}  )
+\frac{ (h_n-h_{n+1})  \Delta_{n-1}  }{  2 J_n  }   (  P_3 -   P_{1}    )
\nonumber \\
h_{n+2}^R && = h_{n+2}+ \Delta_{n+1}  (P_{2+} -P_{2-}  )
-  \frac{ (h_n-h_{n+1})  \Delta_{n+1}  }{  2 J_n  }   (  P_3 -   P_{1}    )
\label{hrb}
\end{eqnarray}

(ii)  the new renormalized coupling between these two spins reads
\begin{eqnarray}
 J_{n-1,n+2}^R && = \frac{ J_{n-1} J_{n+1}  }{   J_n } (P_3+P_1- P_2) 
\label{jrb}
\end{eqnarray}

(iii)  the new renormalized interaction between these two spins reads
\begin{eqnarray}
\Delta^R_{n-1,n+2} &&=  -  \frac{   \Delta_{n-1} \Delta_{n+1} }{  2 J_n  }   (  P_3 -   P_{1}    )
\label{drb}
\end{eqnarray}

This largest-coupling-$J_n$ decimation step has thus for effects to generate smaller renormalized hopping $J^R$
 and smaller interaction $\Delta^R$.  The decimations of the transverse-fields $h_j$ (Eq \ref{omegaa}) and of the couplings $J_j$ (Eq. \ref{omegac})  will thus both drive the system towards the Many-Body-Localized phase.

As a final remark, let us stress that the renormalization rules corresponding to transverse-field decimations (Eqs \ref{hra}, \ref{jra}, \ref{dra})
and to couplings decimations (Eqs \ref{hrb}, \ref{jrb}, \ref{drb}) depend on the values of the emergent LIOMs via the projectors $P_m$,
so that one needs to follow the different branches : the practical implementation of this type of renormalization rules is thus much more involved \cite{rsrgx} than in the ground-state-studies where one focus only on a single branch at each step.
More precisely, for a chain of $N$ quantum spins like Eq. \ref{heisen}, with an Hilbert space of size $2^N$,
the construction of the ground state via the renormalization requires $O(N)$ renormalization steps,
where at each step one systematically chooses the projector associated to the lowest eigenstate of $H_0$ :
 the numerical implementation remains thus elementary and can be applied on very large systems
as a consequence of the polynomial cost with respect to the system size $N$.
However if one wishes to construct the effective dynamics via the RSRG-t or equivalently to construct the whole set
of the $2^N$ eigenstates, one has to follow the whole tree of possibilities for the choice of the projectors
at each renormalization step (see Figure 2a of Ref. \cite{rsrgx}) : since this tree has $2^N$ leaves corresponding to
the $2^N$ eigenstates, the exact numerical implementation of this procedure is limited to small systems as a consequence of
the exponential cost with respect to the size $N$. To overcome this limitation, 
the authors of Ref \cite{rsrgx} have thus proposed to replace the exact application of the renormalization rules on all branches
by a Monte Carlo sampling of the typical branches of the tree (see \cite{rsrgx} for more details and examples of results).
It would be thus interesting to apply the same numerical procedure to the above RG rules
to test their validity in the Many-Body-Localized-Phase by a direct comparison with the true dynamics
as computed via other numerical methods.

\section{Conclusion }

\label{sec_conclusion}

In summary, the Vosk-Altman Strong Disorder Renormalization for the unitary dynamics of random quantum models
has been reformulated within the Floquet theory : the iterative elimination of the local degree of freedom characterized by the highest eigenfrequency $\Omega$ 
can then be taken into account for the neighboring slower degrees of freedom via the two first orders of the high-frequency-expansion of the
effective Floquet Hamiltonian. The output of this reformulation is the general recipe of Eq \ref{veffproj} to derive RG rules.
We have discussed the equivalence with the RSRG-X procedure to construct the whole set of eigenstates.
We have then applied this general framework to the XXZ chain in its Many-Body-Localized phase, by taking into account two possibilities :
the highest eigenfrequency is set either by the biggest transverse field $h_n$ or by the biggest coupling $J_n$. 
More generally, we hope that the present framework will be helpful to analyze the dynamics of other random quantum models.

\end{document}